\documentclass{pinchcr}
\title{Les Houches Lectures on Physics Beyond the Standard Model of Cosmology}
\author{Justin Khoury}
\affiliation{University of Pennsylvania}

\begin{document}

\chapter*{Les Houches Lectures on Physics Beyond the Standard Model of Cosmology}

\centerline{Justin Khoury}

\centerline{ \small{Center for Particle Cosmology, Department of Physics and Astronomy,}}

\centerline{ \small{University of Pennsylvania, Philadelphia, PA 19104}}

\null\hskip5mm

\begin{quote}
In these Lectures, I review various extensions of the $\Lambda$-Cold Dark Matter ($\Lambda$CDM) model, characterized by additional light degrees of freedom in the dark sector. 
In order to reproduce the successful phenomenology of GR in the solar system, these fields must effectively decouple from matter on solar
system/laboratory scales. This is achieved through screening mechanisms, which rely on the interplay between self-interactions and coupling to matter to
suppress deviations from standard gravity. The manifestation of the new degrees of freedom depends sensitively on their
environment, which in turn leads to striking experimental signatures. 
\end{quote}

\chapter{Introduction}

Cosmology can rightfully claim to have a standard model: the $\Lambda$CDM model. With roughly 4.5\% baryonic matter, 26.5\% dark matter,
and 69\% vacuum energy, this model provides an exquisite fit to all known cosmological data. (For a review, see~\shortciteN{Jain:2010ka}.) Its empirical success lies in its parsimony --- only a handful of parameters are required to fit observations. As a result, it is highly predictive: the $\Lambda$CDM expansion and growth histories are tightly correlated, leaving essentially no wiggle room to account for possible discrepancies.

The discomfort with $\Lambda$CDM among theorists is of course one of naturalness. The inferred value of the cosmological constant requires an absurb conspiracy among its various contributions. This is the famous Cosmological Constant (CC) problem. More compelling is the possibility of new physics associated with dark energy, which would stabilize the vacuum energy at its observed value. A parallel with the weak hierarchy problem in particle physics seems appropriate. The recent discovery of the Higgs particle represents the crowning achievement of the Standard Model. The Standard Model is now complete, but leaves us with an unnaturally low value for the Higgs mass. Either the Higgs mass is fine-tuned, or it is stabilized by new physics (such as supersymmetry) at the weak scale. 

The naturalness problem afflicting cosmology is in some sense more robust than its particle physics counterpart. Indeed, the weak hierarchy problem stems from radiative corrections to the Higgs mass from hypothetical particles {\it beyond} the weak scale. In the case of the CC problem, however, vacuum energy contributions from {\it known} particles, such as the electron, are already problematic. Furthermore, the required solution for the CC problem is arguably more radical. While the proposed solutions to the weak hierarchy problem --- supersymmetry, technicolor --- are by no means trivial, they both fit within the standard framework of local quantum field theory.  As shown by~\shortciteN{Weinberg:1988cp}, on the other hand, no dynamical solution to the CC problem is possible within General Relativity (GR).

Let us begin with a lightning review of the CC problem.

\section{The Cosmological Constant Problem}

A precise calculation of the vacuum energy of course requires knowledge of physics all the way to the Planck scale. Within the effective field theory framework,
however, one can calculate robust low-energy contributions to get a sense for the required degree of fine-tuning. 

Consider the vacuum energy of a scalar field of mass $m$, ignoring interactions. Each quantum contributes an energy $E_k = \frac{1}{2} \sqrt{\vec{k}^2 + m^2}$. Integrating over all quanta below some cutoff $\Lambda_{\rm UV}$, the energy density is
\begin{eqnarray}
\nonumber
\langle \rho \rangle &=& \int_0^{\Lambda_{\rm UV}}  \frac{{\rm d}^3k}{(2\pi)^3}  \frac{1}{2} \sqrt{\vec{k}^2 + m^2} \\
\nonumber
&\simeq & \frac{1}{16\pi^2} \left(\Lambda_{\rm UV}^4 + m^2 \Lambda_{\rm UV}^2 - \frac{m^4}{2}  \log\left(\frac{\Lambda_{\rm UV}}{m}\right) + {\rm finite}\right)\,,
\end{eqnarray}
where in the last step we have used the fact that the integral peaks close to $\Lambda_{\rm UV}$, and expanded the integrand in powers of $m/k$. 
The first two terms, which diverge as a power of $\Lambda_{\rm UV}$, depend sensitively on the UV physics. They would be absent, for instance, if one used
instead dimensional regularization. The log term, on the other hand, is regulator-independent and represent a robust, low-energy contribution to the cosmological constant.

The UV-insensitive contribution from all particles with masses $m_i \ll \Lambda_{\rm UV}$ is therefore given by
\begin{equation}
\langle \rho \rangle = -\frac{1}{32\pi^2} \sum_i (-1)^{F_i} m_i^4  \log\left(\frac{\Lambda_{\rm UV}}{m}\right)\,,
\end{equation}
where the sum is over all particle species, and $F_i$ is the fermion number. The electron contribution, for instance, is
\begin{equation}
\rho_{e} \sim m_e^4 \sim 10^{33}~{\rm meV}^4\,,
\end{equation}
which is of course orders of magnitude larger than the observed value of $\sim$meV. This is the essence of the CC problem. The robust, low-energy contributions from {\it known} particles are
unacceptably large. A tremendous cancellation must occur between these contributions and (unknown) high energy contributions to generate the observed value of the vacuum energy. 

An old idea is to postulate that the vacuum energy is effectively time-dependent, and that there is some kind of relaxation mechanism driving the vacuum energy to small values at late times.
Indeed, we need only explain why the vacuum energy is small now, not in the very early universe. Unfortunately, as shown long time ago by~\shortciteN{Weinberg:1988cp}, such a mechanism is impossible,
under the key assumption that gravity is described by GR. A possible loophole is therefore to go beyond GR, which inevitably introduces new degrees of freedom~\shortcite{Feynman:1996kb,Weinberg:1965rz,Deser:1969wk,Khoury:2011ay,Khoury:2013oqa}.

\section{Why screen?}

Although no compelling mechanism has been found to date, we can robustly infer two necessary properties of these degrees of freedom. For concreteness, let us assume they are scalars.
Firstly, to neutralize $\Lambda$ to an accuracy of order $\sim H_0^2M_{\rm Pl}^2 \sim {\rm meV}^4$, the scalars must have a mass at most comparable to $H_0$,
\begin{equation}
m_\phi ~\lower .75ex \hbox{$\sim$} \llap{\raise .27ex \hbox{$<$}} ~H_0\,.
\label{mphi}
\end{equation}
Otherwise, if they were much more massive, they could be integrated out and thus would be irrelevant for the low energy dynamics. 

\begin{figure}[t]
\centering
\includegraphics[width=4in]{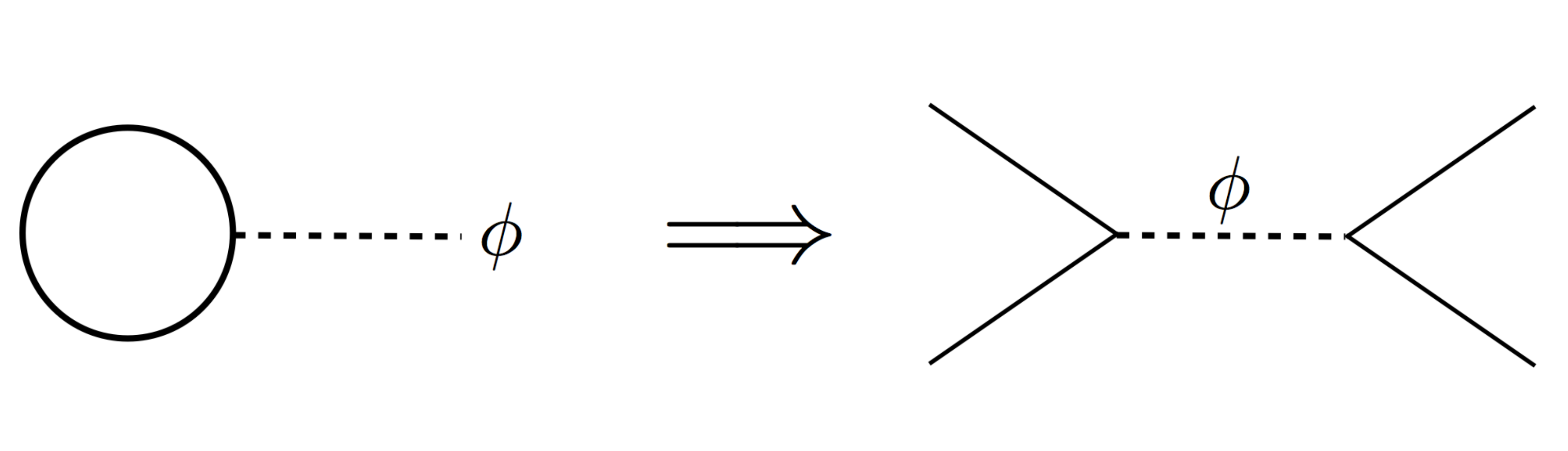}
\caption{\label{CCdiags} \small The tadpole diagram (on the left) involving the scalar field (dotted line) attached to Standard Model fields (solid line) running in the loop is necessary in order to neutralize the Standard Model vacuum energy contribution. By unitarity, the tree-level diagram (on the right) with a scalar exchanged by Standard Model fields is also allowed, implying that the scalar field mediates a 5$^{\rm th}$ force that must therefore be screened. }
\end{figure}

Secondly, these scalars must couple to Standard Model fields, since Standard Model fields contribute $O({\rm TeV}^4)$ to the vacuum energy. In other words, the tadpole diagram shown in Fig.~\ref{CCdiags} must be present in the theory. But by unitarity, so must the exchange diagram. Hence $\phi$ mediates a force between Standard Model fields, whose range, given~(\ref{mphi}), is comparable to the present Hubble radius. If left unabated, such a long-range force would generically lead to significant deviations from General Relativity in the solar system, in conflict with well-known constraints from tests of gravity~\shortcite{Will:2005va}. 

Thus we are led to conclude, on very general grounds, that these scalars must effectively decouple from matter on solar system/laboratory scales in order to reproduce the successful phenomenology of GR. This can be achieved through {\it screening mechanisms}, which rely on the high density of the local environment (relative to the mean cosmological density) to suppress deviations from standard gravity. 
In what follows we will describe 3 general classes of screening mechanisms and highlight some of their characteristic properties.

Besides cosmology and the CC problem, screening mechanisms are also motivated by the vast experimental effort aimed at testing the fundamental nature of gravity on a wide range of scales, from laboratory to solar system to extra-galactic scales. For a review, see~\shortciteN{Will:2005va}. As we will see below, viable screening theories make novel predictions for local gravitational experiments. The subtle nature of these signals have forced experimentalists to rethink the implications of their data and have inspired the design of novel experimental tests. The theories of interest thus offer a rich spectrum of testable predictions for ongoing and near-future tests of gravity.

\chapter{Screening Mechanisms: A Brief Overview}
\label{screenover}

The broad classes of screening mechanisms surveyed here can all be encompassed in the general action
\begin{equation}
L = \frac{M^2_{\rm Pl}}{2}R^{\rm E}  +   L_\phi(\phi,\partial\phi,\partial^2\phi)  + L_{\rm m}\left[A^2(\phi)g_{\mu\nu}^{\rm E}, \psi\right]~,
\label{scalartensoraction}
\end{equation}
where $M_{\rm Pl}^2 = 1/8\pi G_{\rm N}$, and all indices are contracted with the Einstein-frame metric $g_{\mu\nu}^{\rm E}$. For simplicity, we have assumed that the scalar couples conformally and universally to matter fields described by $L_{\rm m}$.\footnote{With chameleons~\shortcite{Khoury:2003aq,Khoury:2003rn}, one can more generally assume different couplings to different matter species, thereby explicitly violating the weak equivalence principle. We also ignored the possibility of disformal coupling, {\it e.g.}~\shortciteN{Koivisto:2012za}.} In particular, matter particles universally follows geodesics of the Jordan-frame metric
\begin{equation}
g_{\mu\nu}^{\rm J} = A^2(\phi)g_{\mu\nu}^{\rm E}\,.
\label{jordan}
\end{equation}

In most situations of interest, save of course for cosmological evolution, we will see that the field excursions are small in Planck units, {\it i.e.} $\Delta\phi \ll M_{\rm Pl}$. (For chameleons, this is true even cosmologically, as shown in~\shortciteN{Wang:2012kj}.) We are therefore justified in linearizing\footnote{An important exception is the symmetron~\shortcite{Hinterbichler:2010es,Olive:2007aj,Pietroni:2005pv,Hinterbichler:2011ca,Brax:2011pk} and varying-dilaton~\shortcite{Brax:2011ja} mechanisms, where a $\phi\rightarrow -\phi$ symmetry precludes the linear term in $A(\phi)$. Instead, $A(\phi) \simeq 1 + g\phi^2/M^2$ in those cases. In practice, the phenomenology of the symmetron is qualitatively similar to that of the chameleon, so for simplicity we focus our discussion to the linear coupling~(\ref{Aphi}).}
the coupling function $A(\phi)$:
\begin{equation}
A(\phi) \simeq 1 + \frac{g \phi}{M_{\rm Pl}}\,.
\label{Aphi}
\end{equation}
The dimensionless coupling $g$ is generally assumed to be $O(1)$, corresponding to gravitational-strength scalar force.
In the Newtonian approximation,~(\ref{jordan}) and~(\ref{Aphi}) imply the following relation between Jordan-frame and Einstein-frame
Newtonian potentials (defined as usual by $g_{00} = -(1+2\Phi)$):
\begin{equation}
\Phi_{\rm J} = \Phi_{\rm E} +  \frac{g \phi}{M_{\rm Pl}}\,.
\label{potrel}
\end{equation}
This will prove useful in physically interpreting the different screening conditions.

For situations relevant to tests of gravity, we can ignore the backreaction of the scalar onto the metric, and approximate $\phi$ as evolving on Minkowski space-time.
Moreover the matter can be treated as a non-relatisvistic source, with negligible pressure. In this approximation, the Lagrangian relevant for the scalar dynamics is
\begin{equation}
L_{\rm scalar} = L_\phi (\phi,\partial\phi,\partial^2\phi) -  \frac{g \phi}{M_{\rm Pl}} \rho_{\rm m}\,,
\label{Lscalarsimple}
\end{equation}
where indices are now contracted with $\eta_{\mu\nu}$. A first requirement on $L_\phi$ is that it must describe a single degree of freedom, {\it i.e.}, its equations of motion must be second-order in time. The most general such Lagrangian (coupled to gravity) was discovered decades ago by~\shortciteN{Horndeski:1974wa}. See~\shortciteN{Deffayet:2011gz} for a recent derivation. A second requirement is that the scalar should develop non-linearities in the presence of sufficiently massive/dense sources, in such a way that its effects are mitigated. It turns out there are three qualitatively different ways to do so, described below.

\begin{itemize}

\vspace{0.2cm}
\item {\bf Chameleon}~\shortcite{Khoury:2003aq,Khoury:2003rn}: 

In this example, the scalar Lagrangian consists of a standard kinetic term plus potential:
\begin{equation}
L_\phi = -\frac{1}{2}(\partial \phi)^2 - V(\phi) \,.
\label{Lcham1}
\end{equation}
Since the interactions are governed by a potential, whether or not the scalar develop non-linearities depends on the local value of $\phi$. In other words, the {\it screening condition} in this case is schematically of the form
\begin{equation}
\phi < \phi_c\,.
\label{chamscreengen}
\end{equation}
We will see in Sec.~\ref{chamsec} that the critical value $\phi_c$ depends on the local gravitational potential. A useful rule of thumb to ascertain which regions of the universe are screened or unscreened is to map out the {\it Newtonian potential} smoothed on some scale. Screened regions correspond to $\Phi_{\rm J} > \Phi_c$. 

\vspace{0.15cm}
A recurring theme of all screening mechanisms is that their strong coupling scale is rather low, {\it i.e.}, comparable or lower than the meV dark energy scale. In the case of chameleons, perturbative unitarity breaks down at the meV scale~\shortcite{Upadhye:2012vh}:
\begin{equation}
\Lambda_{\rm s} \sim {\rm meV}\,.
\end{equation}
A drawback of this screening mechanism is that $L_\phi$ has no particular symmetry, hence one would not expect $V(\phi)$ to be protected under radiative corrections. This issue has been analyzed in detail recently in~\shortciteN{Upadhye:2012vh}, where it was found that keeping quantum corrections under control imposes an upper limit on the chameleon mass in the local (laboratory) environment. The upper limit is very general and insensitive to the details of the potential. Remarkably, for gravitational-strength coupling ($g\sim O(1)$), a factor-of-two improvement over the current experimental bounds on the range of a scalar fifth force would rule out {\it all} viable chameleon models.

\vspace{0.15cm}
On the plus side, since chameleons are described by canonical scalar fields with self-interaction potentials, they in principle admit a standard Wilsonian UV completion. Unlike the other mechanisms described below, they do not suffer from superluminality issues. Ideally one would like to see an explicit UV-complete realization of this mechanism in string theory. As a first step in this direction,~\shortciteN{Hinterbichler:2010wu}
presented a scenario for embedding the chameleon mechanism within supergravity/string theory compactifications. (See~\shortciteNP{Nastase:2013ik,Nastase:2013los} for related work, and~\shortciteNP{Brax:2011qs,Brax:2012mq} for other supersymmetric extensions of chameleon, symmetron, and varying-dilaton theories.) In this approach, the chameleon scalar field is identified with a certain function of the volume modulus of the extra dimensions.  In follow-up work,~\shortciteN{Hinterbichler:2013we} extended the scenario and showed that, with suitable generalization of the superpotential and Kahler potential, the volume modulus can also drive slow-roll inflation in the early universe.

\vspace{0.15cm}
The last generic property of chameleons pertains to their cosmological impact. It was shown recently by~\shortciteNP{Wang:2012kj,Brax:2011aw} that for a general class of chameleon, symmetron and dilaton theories, the range of the scalar-mediated force can be at most $\sim {\rm Mpc}$. Hence, it has negligible effect on density perturbations on linear scales. Nevertheless, the chameleon mechanism
remains interesting as a way to hide light scalars suggested by fundamental theories. The way to test these theories is to study small scale phenomena, as we will see in Sec.~\ref{chamexp}.

\vspace{0.2cm}
\item {\bf Kinetic/K-mouflage}~\shortcite{Babichev:2009ee,Dvali:2010jz}:

In this case, the scalar interactions are governed by a `kinetic' function $P(X)$, where $X  =-(\partial\phi)^2/2$. The prototypical example is
\begin{equation}
L_\phi = - \frac{1}{2} (\partial\phi)^2 - \frac{L^2}{4M_{\rm Pl}^2} (\partial\phi)^3 \,.
\label{kinlag}
\end{equation}
Whether or not $\phi$ becomes non-linear near a source clearly depends on its local gradient, {\it i.e.}, the screening condition in this case takes the form
\begin{equation}
|\partial \phi| > L^{-1} M_{\rm Pl}\,.
\end{equation}
In light of~(\ref{potrel}), the screening condition can equivalently be cast as a condition on the local gravitational acceleration, $|\nabla \Phi_{\rm J}| > |\vec{a}_c|$. A useful rule of thumb to ascertain
which regions of the universe are screened or unscreened is to map out the {\it gravitational acceleration} smoothed on some scale. Cosmologically, the departures from standard gravity are most pronounced on scales $ \lower .75ex \hbox{$\sim$} \llap{\raise .27ex \hbox{$>$}} ~ {\rm Mpc}$, that is, on linear scales. 

\vspace{0.15cm}
Choosing $L \sim H_0$, the strong coupling scale is conveniently of the order of the dark energy scale:
\begin{equation}
\Lambda_{\rm s} \sim \sqrt{L^{-1}M_{\rm Pl}} \sim {\rm meV}\,.
\end{equation}
The scalar field enjoys a shift symmetry 
\begin{equation}
\phi\rightarrow \phi + c\,. 
\end{equation}
%s
This is mildly broken by the coupling to matter, but given the low value of the cutoff, loop corrections to the mass are small, {\it i.e.},
$\delta m _\phi \sim \Lambda_{\rm s}^2/M_{\rm Pl} \sim H_0$, which is potentially interesting for late-time cosmology.

\vspace{0.15cm}
Screening only works for the particular sign of the $(\partial\phi)^4$ term given in~(\ref{kinlag}). Indeed, assuming spherical symmetry, the equation of motion outside a point source of mass $M$ reduces to, upon using Gauss' law,
\begin{equation}
 \frac{{\rm d}\phi}{{\rm d} r}\bigg(1    + \frac{L^2}{M_{\rm Pl}^2} \left( \frac{{\rm d}\phi}{{\rm d} r}\right)^2\bigg)= \frac{g M}{4\pi r^2 M_{\rm Pl}}\,.
\end{equation}
Screening relies on the non-linear term dominating close to the source. Very far from the source, on the other hand, the equation is to a good approximation linear. Only for this particular choice of sign can the linear and non-linear regimes match continously. Unfortunately, this choice of sign implies that radial perturbations on top of this background propagate superluminally. Correspondingly, the $2\rightarrow 2$ amplitude computed from~(\ref{kinlag}) fails to satisfy the standard analyticity properties of local quantum field theories~\shortcite{Adams:2006sv}.

\vspace{0.2cm}
\item {\bf Vainshtein}~\shortcite{Vainshtein:1972sx,ArkaniHamed:2002sp,Deffayet:2001uk}:

In this case, the scalar interactions involve second-derivatives of $\phi$. The prototypical Lagrangian is the so-called cubic Galileon~\shortcite{Deffayet:2001uk,Luty:2003vm}:
\begin{equation}
L_\phi = - \frac{1}{2} (\partial\phi)^2 - \frac{L^2}{M_{\rm Pl}} \Box\phi (\partial\phi)^2\,.
\label{Lgal}
\end{equation}
Whether or not $\phi$ develops non-linearities near a massive source, that is, whether or not the cubic term is large compared to the kinetic term, depends on the magnitude of $\Box\phi$.
The screening condition is
\begin{equation}
|\partial^2 \phi| > L^{-2} M_{\rm Pl}\,.
\end{equation}
In light of~(\ref{potrel}), this is equivalent to a condition on the local curvature, $|\nabla^2 \Phi_{\rm J}| > |R_c|$. A useful rule of thumb to ascertain
which regions of the universe are screened or unscreened is to map out the {\it curvature} smoothed on some scale. Like the kinetic mechanism,
the departures from standard gravity are most significant today on scales $ \lower .75ex \hbox{$\sim$} \llap{\raise .27ex \hbox{$>$}} ~ {\rm Mpc}$. 

\vspace{0.15cm}
Choosing $L \sim H_0$, the strong coupling scale is 
\begin{equation}
\Lambda_{\rm s} \sim \sqrt{L^{-2}M_{\rm Pl}} \sim (1000~{\rm km})^{-1}\,.
\label{Lsrel}
\end{equation}
As we will see in Sec.~\ref{galsec}, however, the strong coupling scale is renormalized to much higher values in the vicinity of massive objects, such as in the solar system.  

\vspace{0.15cm}
The scalar field enjoys a galilean-like internal symmetry
\begin{equation}
\phi\rightarrow \phi + c + b_\mu x^\mu\,,
\end{equation}
where $b_\mu$ is an arbitrary constant 4-vector. Indeed, under this transformation both terms in~(\ref{Lgal}) shift by a total derivative. 
This is mildly broken by the coupling to matter, but the low cutoff value implies a tiny quantum correction to the mass $\delta m _\phi \sim \Lambda_{\rm s}^2/M_{\rm Pl} \ll H_0$.

This theory also suffers from superluminal propagation around certain backgrounds. In particular, radial perturbations around a spherically-symmetric source propagate
strictly superluminally, as we will see explicitly in Sec.~\ref{galsec}. Nevertheless, it has been argued that galileons are protected against the formation of closed time-like curves (CTC) by a 
Chronology Protection~\shortcite{Burrage:2011cr,Evslin:2011rj}, analogously to what happens in GR~\shortcite{Hawking:1991nk}. Specifically, if one tries to create a closed causal time-like curve from healthy initial conditions,
the galileon will become strongly coupled ({\it i.e.}, the effective description will break down) before the CTC can form.

\end{itemize}

One may wonder whether there are other possibilities, such as screening conditions involving higher-derivatives of the field, {\it e.g.} $|\partial^3 \phi | > M^4$? 
The answer is no, at least in the context of Lorentz-invariant theories. The Lagrangian in this case would include terms with $n\geq 3$ derivatives per field, which would necessarily lead to higher-order
equations of motion, and hence ghost instabilities. The 3 classes listed above, with the screening condition involving $\phi$, $\partial\phi$ and $\partial^2\phi$,
are the only possibilities. Of course one could always consider hybrid versions of these mechanisms. 

\begin{table}[t]
\tableparts
{
\caption{Classification of screening mechanisms and their properties.}
\label{sample table}
}
{
\begin{tabular}{l c l l c l l c l l c l | c | | c |}
\hline
 & & & & & & &  \\[-6pt]
Mechanism & Screening Condition & Symmetry & Cutoff & Superluminality & Scales  \\[3pt]
\hline
\hline 
 & & & & & & &  \\[-6pt]
Chameleon        & $\phi < \phi_c$ & None & $\Lambda_{\rm s} \sim {\rm mm}^{-1}$ & No & $\lower .75ex \hbox{$\sim$} \llap{\raise .27ex \hbox{$<$}} ~ {\rm Mpc}$  \\[3pt]
\hline
& & & & & & &  \\[-6pt]
Kinetic & $|\partial\phi| > \Lambda_{\rm s}^2$ & $\delta\phi = c$ &  $\Lambda_{\rm s} \sim {\rm mm}^{-1}$  & Yes & $ \lower .75ex \hbox{$\sim$} \llap{\raise .27ex \hbox{$>$}} ~ {\rm Mpc}$ \\[3pt]
\hline
 & & & & & & &  \\[-6pt]
Vainshtein  & $|\partial^2\phi| > \Lambda_{\rm s}^3$ & $\delta\phi = c + b_\mu x^\mu$ &  $\Lambda_{\rm s} \sim (1000~{\rm km})^{-1}$  & Yes & $ \lower .75ex \hbox{$\sim$} \llap{\raise .27ex \hbox{$>$}} ~ {\rm Mpc}$  \\[3pt]
\hline
\end{tabular}
}
\end{table}

\chapter{Chameleons}
\label{chamsec}

The chameleon mechanism operates whenever a scalar field couples to matter in such a way that its effective mass depends on the local matter density~\shortcite{Khoury:2003aq,Khoury:2003rn,Gubser:2004uf,Brax:2004qh,Mota:2006ed,Mota:2006fz}. The scalar-mediated force between matter particles can be of gravitational strength, but its range is a decreasing function of ambient matter density, thereby avoiding detection in regions of high density. Deep in space, where the mass density is low, the scalar is light and mediates a fifth force of gravitational strength, but near the Earth, where experiments are performed, and where the local density is high, it acquires a large mass, making its effects short range and hence unobservable. 

This is achieved with a canonical scalar field with suitable self-interaction potential $V(\phi)$. The theory (in the weak-field limit and for non-relativistic matter) is given by~(\ref{Lscalarsimple}) and~(\ref{Lcham1}):
\begin{equation}
L_{\rm cham} =  -\frac{1}{2}(\partial \phi)^2 - V(\phi) -  \frac{g \phi}{M_{\rm Pl}} \rho_{\rm m}\,.
\end{equation}
The dimensionless coupling parameter $g$ is assumed to be~$O(1)$, corresponding to gravitational strength coupling. 
The equation of motion for $\phi$ that derives from this Lagrangian is
\begin{equation}
\nabla^2 \phi = V_{,\phi} +  \frac{g}{M_{\rm Pl}} \rho_{\rm m} \,.
\label{phigen}
\end{equation}
The immediate thing to notice is that, because of its coupling to matter, the scalar field is governed by a density-dependent effective potential
\begin{equation}
V_{\rm eff}(\phi) = V(\phi) + \frac{g \phi}{M_{\rm Pl}} \rho_{\rm m} \,.
\label{Veffcham}
\end{equation}

\begin{figure}[t]
\centering
\includegraphics[width=3.1in]{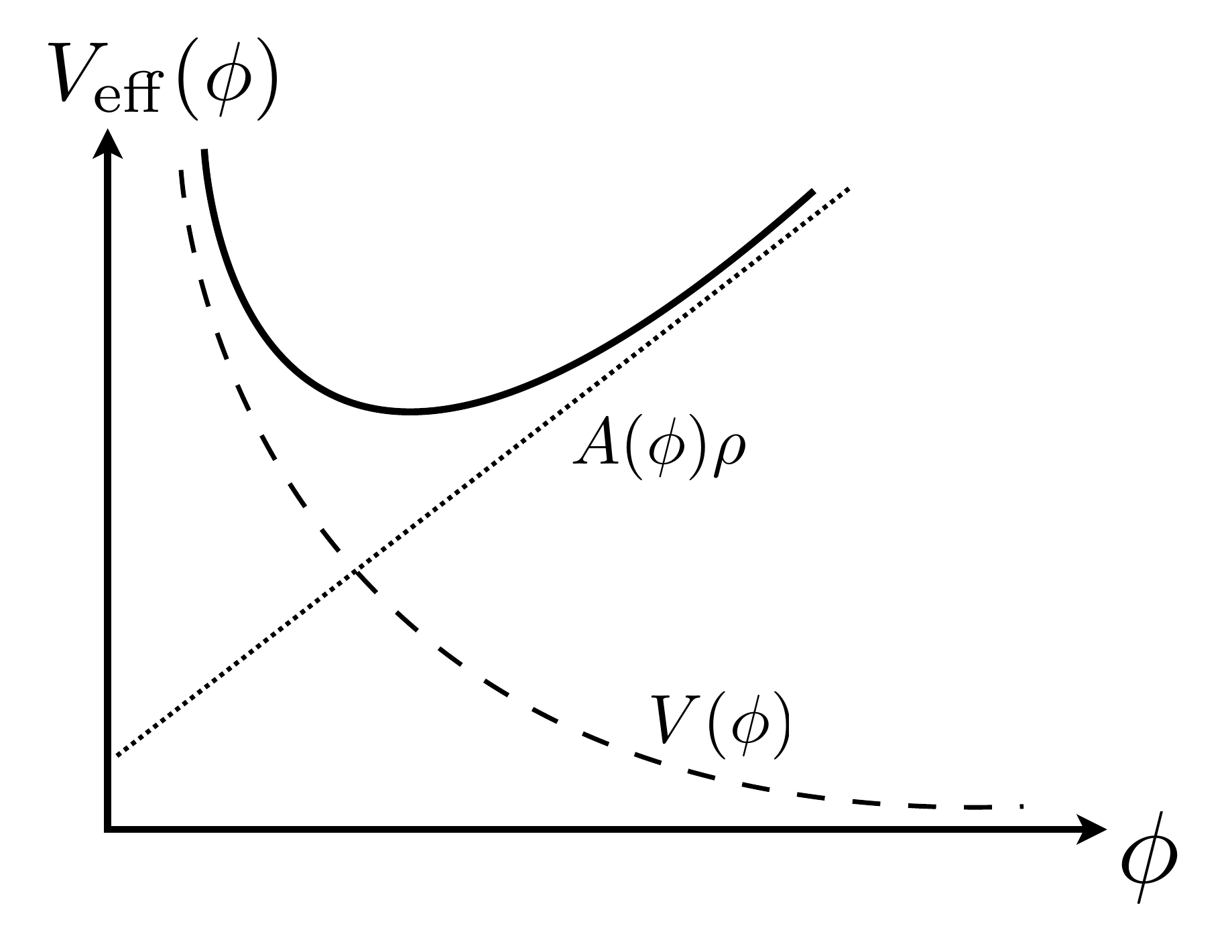}
\caption{\label{chameffpot} \small Schematic of the effective potential felt by a chameleon field (solid line). The effective potential is a sum of the bare potential of runaway form, $V(\phi)$ (dashed line), and a density-dependent piece, from coupling to matter (dotted line).}
\end{figure}

\begin{figure}[t]
\centering
\includegraphics[width=5.1in]{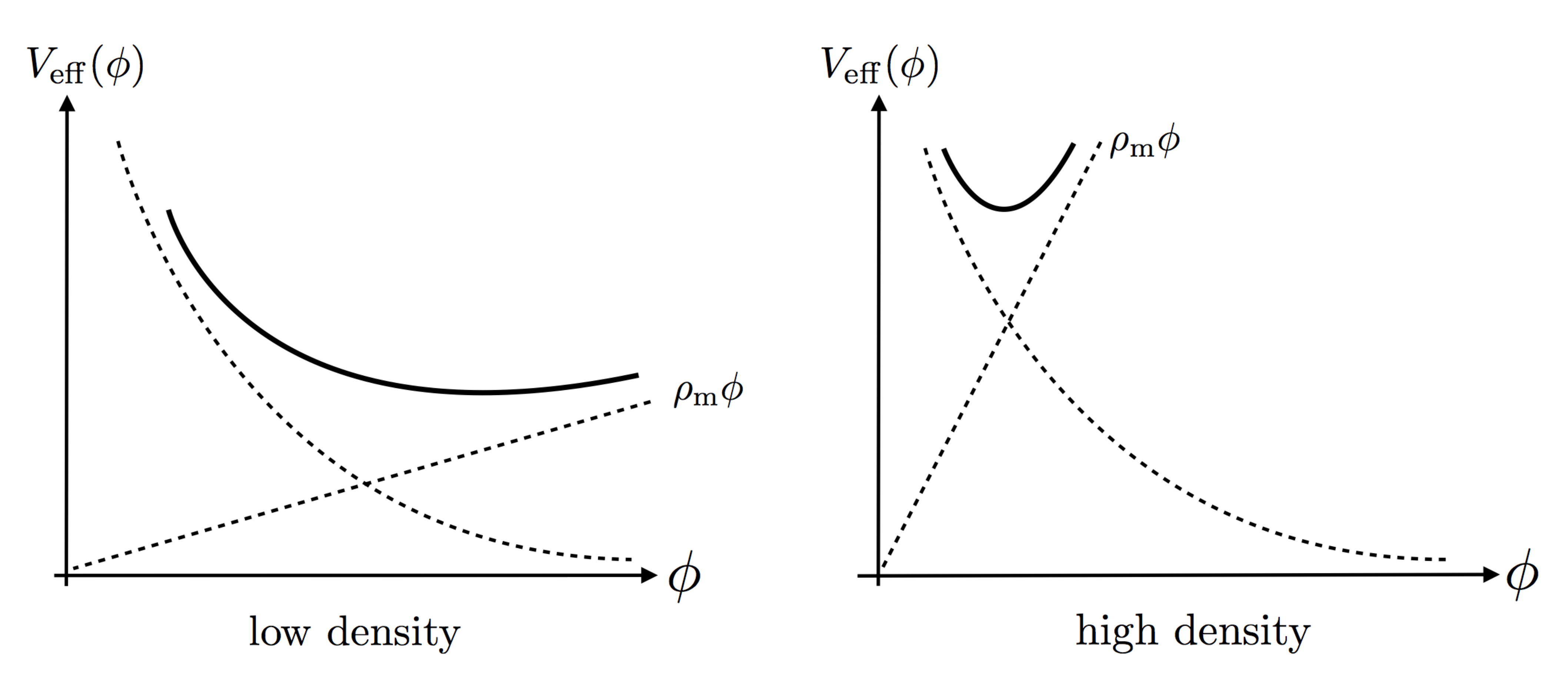}
\caption{\label{champotcomp} \small Effective potential for $a)$ low ambient matter density  and $b)$ high ambient density. As the density increases, the minimum of the effective potential, $\phi_{\rm min}$, shifts to smaller values, while the mass of small fluctuations, $m_\phi$, increases.}
\end{figure}

For suitably chosen $V(\phi)$, the effective potential can develop a minimum  at some finite field value $\phi_{\rm min}$ in the presence of background matter density.
This is illustrated in Fig.~\ref{chameffpot} for a monotonically-decreasing $V(\phi)$ and monotonically-increasing $A(\phi)$. As shown in Fig.~\ref{champotcomp}, the
mass of small fluctuations around the minimum,
\begin{equation}
m^2_\phi  = V_{,\phi\phi} (\phi_{\rm min})\,,
\end{equation}
{\it increases}, while $\phi_{\rm min}$ decreases, with increasing density. 

A prototypical chameleon potential satisfying these properties is the inverse power-law form, 
\begin{equation}
V(\phi) = \frac{\Lambda^{4+n}}{\phi^n}\,.
\label{trackerpot}
\end{equation}
(Potentials with {\it positive} powers of the field, $V(\phi) \sim \phi^{2s}$ with $s$ an integer $\geq 2$, are also good candidates for chameleon theories~\shortcite{Gubser:2004uf}.) 
The effective potential in this case is given by
\begin{equation}
V_{\rm eff}(\phi) = \frac{\Lambda^{4+n}}{\phi^n} + \frac{g \phi}{M_{\rm Pl}} \rho_{\rm m} \,.
\end{equation}
For $g > 0$, this displays a minimum at 
\begin{equation}
\phi_{\rm min} = \left( \frac{n\Lambda^{4+n}M_{\rm Pl}}{g\rho_{\rm m}}\right)^{\frac{1}{n+1}}\,.
\end{equation}
The mass of small fluctuations around the minimum is
\begin{equation}
m^2_\phi  = \frac{n(n+1)\Lambda^{4+n}}{\phi_{\rm min}^{n+2}} \sim \rho_{\rm m}^{\frac{n+2}{n+1}}\,.
\label{mrho}
\end{equation}
Thus $\phi_{\rm min}$ and $m_\phi^2$ are respectively decreasing and increasing functions of  the background density $\rho_{\rm m}$, as desired. 

The tightest constraint on the model comes from laboratory tests of the inverse square law, which set an upper limit of $\approx 40\;\mu$m on the fifth-force range assuming gravitational strength coupling~\shortcite{Adelberger:2006dh}. This imposes the following bound at laboratory density $\rho_{\rm lab} \sim 10^{-3}~{\rm g}/{\rm cm}^3$:
\begin{equation}
m_{\phi}^{-1} (\rho_{\rm lab}) \; \lower .75ex \hbox{$\sim$} \llap{\raise .27ex \hbox{$<$}}\;  40~\mu {\rm m}\,.
\label{labcons}
\end{equation}
Plugging in numbers for the inverse power-law potential, this translates to an upper bound on the scale $M$~\shortcite{Khoury:2003aq,Khoury:2003rn}
\begin{equation}
\Lambda\; \lower .75ex \hbox{$\sim$} \llap{\raise .27ex \hbox{$<$}}\; {\rm meV}\,,
\label{Mlim}
\end{equation}
which, remarkably, coincides with the dark energy scale. (There is some mild dependence on $n$, which we ignore for simplicity.)
This also ensures consistency with all known constraints on deviations from GR, including post-Newtonian tests in the solar system and binary pulsar observations~\shortcite{Khoury:2003aq,Khoury:2003rn}. Equation~(\ref{labcons}) is at best a rough guess, however, since tests of the inverse square law are performed in vacuum chambers. A careful modeling of the E$\ddot{{\rm o}}$t-Wash set-up was done in~\shortciteNP{Khoury:2003aq,Khoury:2003rn}, including a calculation of the chameleon profile inside the vacuum chamber. The end result is that Ref.~(\ref{Mlim}) is a fairly accurate bound. 

Using~(\ref{mrho}), the bound~(\ref{labcons}) translates to the following range for cosmic density $\rho_{\rm cosmos} \sim 10^{-30}~{\rm g}/{\rm cm}^3$:
\begin{equation}
m_{\rm phi}^{-1} (\rho_{\rm cosmos})  \; \lower .75ex \hbox{$\sim$} \llap{\raise .27ex \hbox{$<$}}\;  {\rm Mpc}\,.
\end{equation}
(This bound is achieved assuming for $n\ll 1$. For $n\gg 1$, on the other hand, the upper bound becomes $10^{-7}~{\rm pc}$.) The Compton wavelength of the chameleon cosmologically is at least a factor of $10^3$ shorter than the Hubble radius $H_0^{-1}$, which would be a desirable value for quintessence-like behavior. In particular, chameleon effects are Yukawa-suppressed, and GR is recovered, above the Mpc scale. This confirms the claim made in Sec.~\ref{screenover} that chameleons have negligible impact on linear-scale density perturbations today. Although derived above assuming an inverse power-law potential, the Mpc barrier is actually more general. As shown in~\shortciteN{Wang:2012kj}, this holds under very general conditions whenever the screening condition is determined by the local $\phi$ value, as is the case for chameleon, symmetron and varying-dilaton scalar fields.

Nevertheless the range of the chameleon-mediated force spans a remarkable factor of $10^{27}$ from the laboratory to the cosmos! The scalar field effectively uses the exponentially large density of the local environment (compared to cosmological density) to hide itself from experiments, which is why it aptly deserves the name ``chameleon".

At a typical place in the solar system, the matter density in baryonic gas and dark matter is approximately $\rho_{{\rm solar}-{\rm system}} \sim 10^{-25}~{\rm g}/{\rm cm}^3$. 
The corresponding Compton wavelength for the chameleon is
\begin{equation}
m_{\phi}^{-1} (\rho_{{\rm solar}-{\rm system}} ) \; \lower .75ex \hbox{$\sim$} \llap{\raise .27ex \hbox{$<$}}\;  10^6~{\rm AU}\,,
\end{equation}
which is orders of magnitude larger than the size of the solar system! In other words, the chameleon mediates a long-range force in the solar system, which should have been seen
in post-Newtonian tests of GR. Fortunately, we will see that the chameleon has another way of hiding itself, by suppressing its coupling to massive objects, such as the Sun
or the Earth. This is the so-called {\it thin-shell} effect, which we describe next. 

\section{Spherically symmetric source and thin-shell effect}

In order to understand in detail how the chameleon force is suppressed in the presence of high ambient density, we want to solve the field profile in the presence of a massive compact object, following~\shortciteN{Khoury:2003rn}. We consider a spherically symmetric object with radius $R$, homogeneous density $\rho_{\rm in}$ and total mass $M$. Further, we imagine that this object is immersed in a homogeneous medium with density $\rho_{\rm out}$. In the case of the Sun or the Earth, this is meant to model the density of the ambient baryonic gas and dark matter.

\begin{figure}[t]
\centering
\includegraphics[width=5.1in]{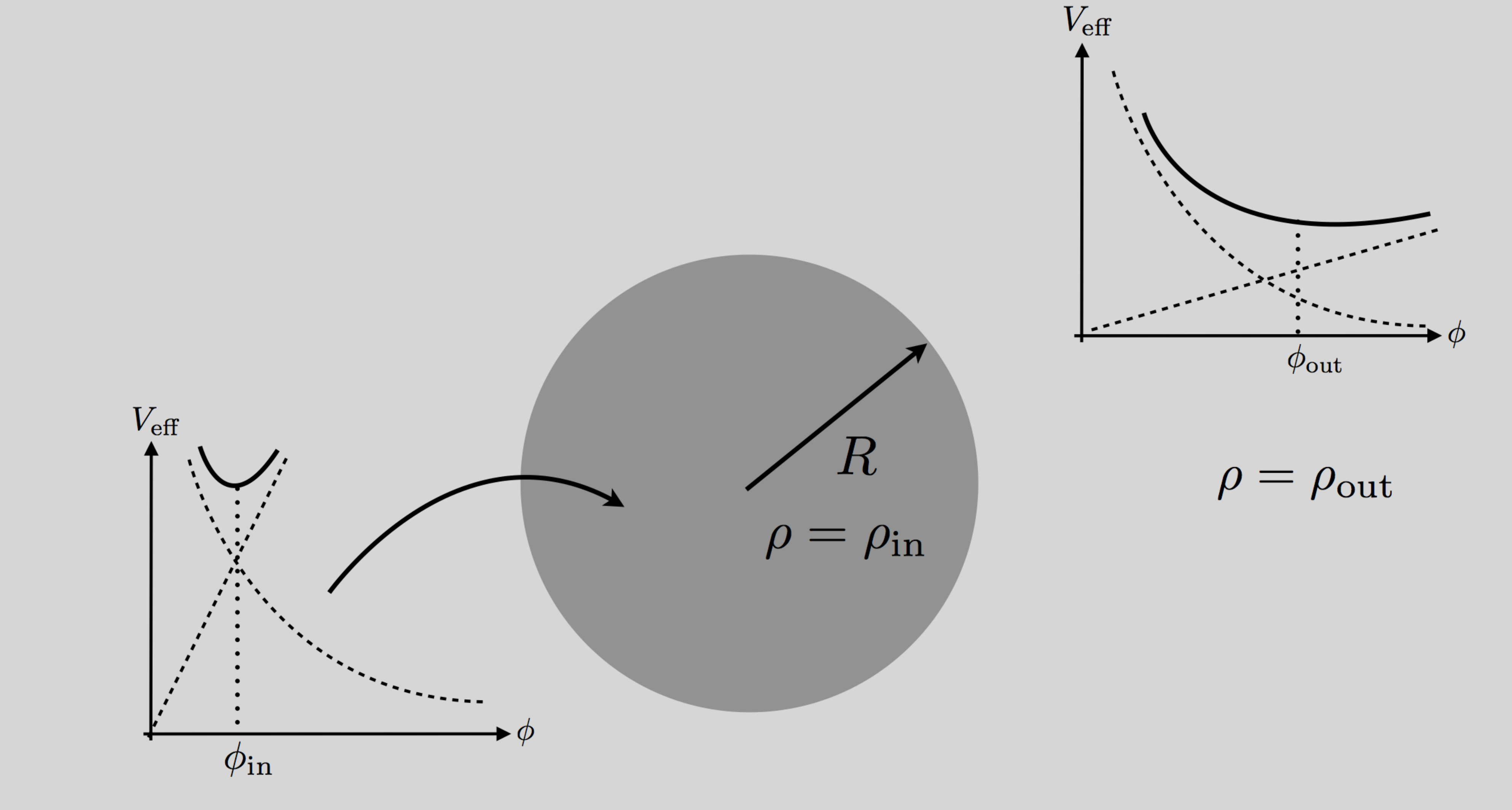}
\caption{\label{chambody} \small Sketch of the set-up for thin-shell calculation.}
\end{figure}

The scalar equation of motion~(\ref{phigen}) for a static and spherically-symmetric background reduces to
\begin{equation}
\frac{{\rm d}^2\phi}{{\rm d}r^2} +\frac{2}{r}\frac{{\rm d}\phi}{{\rm d}r} = V_{,\phi}  + g \frac{\rho}{M_{\rm Pl}}~,
\label{chameombody}
\end{equation}
where the density profile is given by
\begin{equation}
\rho_{\rm m}(r) = \left\{\begin{array}{l}
\rho_{\rm in}~;~~~~~~~~r < R
\\
\rho_{\rm out}~;~~~~~~~r > R
\end{array}\right.~,
\end{equation}
with corresponding minima of the effective potential denoted by $\phi_{\rm min}$ and $\phi_{\rm out}$, respectively. The corresponding mass $m_\phi$ around these minima will be similarly denoted.

The situation is sketched in Fig.~\ref{chambody}. As boundary conditions, the field profile must be regular at the origin, ${\rm d}\phi/{\rm d} r = 0$ at $r=0$, and minimize the effective potential far from the source,
$\phi\rightarrow \phi_{\rm out}$ as $r\to \infty$.

It is instructive to derive the general solution through simple analytical arguments~\shortcite{Khoury:2003aq,Khoury:2003rn}. For a sufficiently large body, in a sense that will be made precise below, the field approaches the minimum of its effective potential deep in the interior:
\begin{equation}
\phi \simeq \phi_{\rm in}\,;\qquad r < R \,.
\label{phiin}
\end{equation}
Outside the object, but still within an ambient Compton wavelength away ($ r < m^{-1}_{\rm out}$), the field profile is approximately $1/r$:
\begin{equation}
\phi \simeq \frac{C}{r} +  \phi_{\rm out} \,;\qquad R < r < m^{-1}_{\rm out}\,,
\label{phiout}
\end{equation}
where we have imposed the boundary condition $\phi\rightarrow \phi_{\rm out}$ as $r\to \infty$. The constant $C$ is fixed by matching~(\ref{phiin}) and~(\ref{phiout})
at the surface of the object, {\it i.e.}, $\phi(r = R) = \phi_{\rm in}$. As a result, the exterior solution is
\begin{equation}
\phi \simeq - \frac{R}{r} (\phi_{\rm out} - \phi_{\rm in}) + \phi_{\rm out}\,.
\label{phiext}
\end{equation}

The above solution admits a nice analogy with electrostatics~\shortcite{JonesSmith:2011tn,Pourhasan:2011sm}. Indeed, since $\nabla^2 \phi\simeq 0$ both inside and outside the source, the body acts as a conducting sphere. Any chameleon charge is confined to a thin shell of thickness $\Delta R$ near the surface. The surface ``charge density" $g \rho_{\rm in} \Delta R/M_{\rm Pl}$ supports the discontinuity in field gradients:
\begin{equation}
\left.\frac{{\rm d} \phi}{{\rm d}r}\right\vert_{r=R^+} = \frac{g \rho_{\rm in}}{M_{\rm Pl}} \Delta R\,.
\end{equation}
Substituting~(\ref{phiext}), we can solve for the shell thickness:
\begin{equation}
\frac{\Delta R}{R} = \frac{\phi_{\rm out} - \phi_{\rm in}}{6g M_{\rm Pl}\Phi} \,,
\end{equation}
where $\Phi \equiv \frac{M}{8\pi M_{\rm Pl}^2R}$ is the surface gravitational potential. For consistency, we should have $\Delta R  \ll R$. An object is therefore said to be {\it screened} if
\begin{equation}
\frac{\phi_{\rm out} - \phi_{\rm in}}{6g M_{\rm Pl}\Phi} \ll 1\,.
\label{chamscreencond}
\end{equation}
The exterior field profile in this case can be written as
\begin{equation}
\phi(r> R) \simeq  -\frac{g}{4\pi M_{\rm Pl}} \frac{3\Delta R}{R} \frac{Me^{-m_{\rm out}(r-R)}}{r} \qquad ({\rm screened})\,,
\end{equation}
where we have restored the Yukawa exponential. This profile is identical to that of a massive scalar of mass $m_{\rm out}$, except
that the coupling is reduced by the thin-shell factor $\Delta R/R \ll 1$.

Clearly the above approximations break down if $\frac{\phi_{\rm out} - \phi_{\rm in}}{6g M_{\rm Pl}\Phi}\, \lower .75ex \hbox{$\sim$} \llap{\raise .27ex \hbox{$>$}}\, 1 $. For fixed density contrast, this
corresponds to a weak source ({\it i.e.}, one with small $\Phi$). In this regime, the coupling is not suppressed, and the object is said to be {\it unscreened}:
\begin{equation}
\phi (r>R) \simeq  -\frac{g}{4\pi M_{\rm Pl}} \frac{Me^{-m_{\rm out}(r-R)}}{r} \qquad ({\rm unscreened})\,.
\end{equation}
This is the usual Yukawa profile for a massive scalar. 

This chameleon screening effect can also be understood qualitatively as follows. If the object is sufficiently massive such that deep inside the object the chameleon minimizes 
the effective potential for the interior density, then the mass of chameleon fluctuations is relatively large inside the object. As a result, the contribution from the core to the exterior
profile is Yukawa-suppressed. Only the contribution from within a thin shell beneath the surface contributes significantly to the exterior profile. In other words, since the chameleon effectively couples
only to the shell, whereas gravity of course couples to the entire bulk of the object, the chameleon force on an exterior test mass is suppressed compared to the gravitational force.

\section{Experimental/Observational Tests}
\label{chamexp}

The idea that the manifestation of a fifth force is sensitive to the environment has spurred a lot of activity. Astrophysically, chameleon scalars affect the
internal dynamics~\shortcite{Hui:2009kc,Jain:2011ji} and stellar evolution~\shortcite{Chang:2010xh,Davis:2011qf,Jain:2012tn} of dwarf galaxies residing in void or
mildly overdense regions. In the laboratory, chameleons have motivated multiple experimental efforts aimed at searching for their signatures:

\vspace{0.1cm}
\noindent $\bullet$ The E\"ot-Wash experiment searches for deviations from the inverse-square-law at distances $\mathrel{\mathstrut\smash{\ooalign{\raise2.5pt\hbox{$>$}\cr\lower2.5pt\hbox{$\sim$}}}} 40\;\mu$m. Based on detailed theoretical predictions~\shortcite{Upadhye:2006vi}, the E\"ot-Wash group analyzed their data to constrain part of the chameleon parameter space~\shortcite{Adelberger:2006dh}.

\vspace{0.1cm}
\noindent $\bullet$ If chameleons interact with the electromagnetic field via $e^{\beta_\gamma\phi}F_{\mu\nu}F^{\mu\nu}$, then photons traveling in a magnetic field will undergo photon/chameleon oscillations, similar to axions. The CHameleon Afterglow SEarch (CHASE) experiment~\shortcite{Chou:2008gr,Steffen:2010ze} has looked for an afterglow from trapped chameleons converting into photons. Similarly, the  Axion Dark Matter eXperiment (ADMX) resonant microwave cavity was used to search for chameleons~\shortcite{Rybka:2010ah}. Photon-chameleon mixing can occur deep inside the Sun~\shortcite{Brax:2010xq} and affect the spectrum of distant astrophysical objects~\shortcite{Burrage:2008ii}. 

Through the nice analogy between chameleon screening and electrostatics~\shortcite{JonesSmith:2011tn,Pourhasan:2011sm} mentioned above, a novel experimental test of chameleons has been proposed
recently that would exploit an enhancement of the scalar field near the tip of pointy objects --- a``lightning rod" effect~\shortcite{JonesSmith:2011tn}. See~\shortciteN{Brax:2009ey} for a
discussion of collider signatures, and~\shortciteN{Upadhye:2012rc} for signatures of symmetron in the laboratory.

The most striking signature of chameleons can be found by testing gravity in space. Because the screening condition depends on the ambient density, small bodies that are screened in the
laboratory may be unscreened in space. This leads to striking predictions for future satellite tests of gravity, such as the planned MicroSCOPE mission\footnote{{\tt http://microscope.onera.fr/}} and STE-QUEST\footnote{{\tt http://sci.esa.int/science-e/www/area/index.cfm?fareaid=127}}. 
In particular, chameleons can result in violations of the (weak) Equivalence Principle in orbit with $\eta \equiv \Delta a/a \gg 10^{-13}$, in blatant conflict with laboratory constraints. Similarly,
the total force --- gravitational + chameleon-mediated --- between unscreened particles can be $O(1)$ larger than in standard gravity, which would appear as $O(1)$ deviations
from the value of $G_{\rm N}$ measured on Earth.

\chapter{Galileons}
\label{galsec}

Galileons are scalar field theories with a number of remarkable properties. The simplest galileon theory~\shortcite{Deffayet:2001uk,Luty:2003vm,Nicolis:2004qq} was
originally discovered as a particular decoupling limit of the Dvali-Gabadadze-Porrati model~\shortcite{Dvali:2000hr,Dvali:2000xg}. This was generalized by~\shortciteN{Nicolis:2008in}
to enumerate all possible galileon terms.\footnote{Historically, these scalar field theories were actually discovered much earlier~\shortcite{Horndeski:1974wa,Fairlie:1992zn,Fairlie:1992nb,Fairlie:1991qe}.}
Galileons also appear in the decoupling limit of the de Rham-Gabadadze-Tolley (dRGT) massive gravity theories~\shortcite{deRham:2010kj,deRham:2010kj,Hassan:2011hr,Hinterbichler:2011tt} and its extensions~\shortcite{D'Amico:2012zv,Gabadadze:2012tr}, as well as cascading gravity~\shortcite{deRham:2007xp,deRham:2007rw,deRham:2009wb,deRham:2010rw,Agarwal:2009gy,Agarwal:2011mg}.

Galileons have been covariantized~\shortcite{Deffayet:2009wt,Deffayet:2009mn}, used to drive cosmic acceleration~\shortcite{Chow:2009fm,Silva:2009km,DeFelice:2010as,Deffayet:2010qz}, applied to inflation~\shortcite{Burrage:2010cu,Creminelli:2010qf,Mizuno:2010ag,DeFelice:2011zh,Kobayashi:2011pc,RenauxPetel:2011dv,Gao:2011qe,RenauxPetel:2011uk}, and employed as a self-tuning mechanism~\shortcite{Charmousis:2011bf,Copeland:2012qf}. They have inspired novel scenarios of the early universe, specifically Galilean Genesis and related scenarios~\shortcite{Creminelli:2010ba,Hinterbichler:2011qk,LevasseurPerreault:2011mw,Liu:2011ns,Qiu:2011cy,Wang:2012bq,Liu:2012ww,Hinterbichler:2012mv,Creminelli:2012my,Creminelli:2012qr,Hinterbichler:2012fr,Hinterbichler:2012yn}.

Galileons have been extended to $p$-forms~\shortcite{Deffayet:2010zh}, multiple fields~\shortcite{Padilla:2010de,Padilla:2010ir,Padilla:2010tj,Hinterbichler:2010xn},
and to more general backgrounds~\shortcite{Goon:2011qf,Burrage:2011bt,Goon:2011uw,Goon:2011xf}. See~\shortciteNP{Khoury:2010gb,Khoury:2011da,Koehn:2012ar,Koehn:2012te,Koehn:2013hk} for recent embeddings of galileons and general higher-derivative scalar theories in supersymmetry and supergravity.

At the heart of the phenomenological viability of these theories is the Vainshtein screening mechanism~\shortcite{Vainshtein:1972sx,ArkaniHamed:2002sp,Deffayet:2001uk}.
This relies on derivative couplings of a scalar field becoming large in the vicinity of massive sources. These non-linearities crank up the kinetic term of
perturbations, thereby weakening their interactions with matter. 

\section{Galileon Basics}

The defining properties of galileons are: $i)$ their equations of motion are second-order; $ii)$ they enjoy the ``galilean" shift symmetry
\begin{equation}
\pi   \rightarrow \pi + c + b_\mu x^\mu\,.
\label{galsym}
\end{equation}
The simplest, non-trivial theory with these properties is the cubic galileon theory given in Eq.~(\ref{Lgal}):
\begin{equation}
L_{\rm gal} =  -3 (\partial \pi)^2 - \frac{1}{\Lambda_{\rm s}^3} \Box\pi (\partial\pi)^2  -  \frac{g}{M_{\rm Pl}} \pi \rho_{\rm m}\,.
\label{L3}
\end{equation}
(To conform to the literature, we denote the galileon scalar by $\pi$, with appropriately rescaled its kinetic term.)  
To check property $ii)$, it is clear that, apart from the coupling to matter, the scalar terms are strictly invariant under the ordinary shift symmetry $\delta\pi = c$. 
They also shift by a total derivative under $\delta\pi =  b_\mu x^\mu$, as can be checked explicitly:
\begin{eqnarray}
\nonumber
\delta(\partial\pi)^2 &=& 2b^\mu \partial_\mu\pi  = 2\partial_\mu(b^\mu\pi)\,;\\
\delta \left((\partial\pi)^2\Box\pi\right) &=& 2b^\mu\partial_\mu \pi\Box\pi = - 2b^\mu\partial_\mu\partial_\nu \pi \partial^\nu\pi = -\partial_\mu \left(b^\mu(\partial\pi)^2\right)\,,
\end{eqnarray}
where we have integrated by parts a bunch of times. Hence this is a symmetry of the action.

To check property $i)$, we vary~(\ref{L3}) to obtain the equation of motion:
\begin{equation}
3\Box\pi + \frac{1}{\Lambda_{\rm s}^3}\bigg( (\Box\pi)^2 - (\partial_\mu\partial_\nu\pi)^2\bigg) = \frac{g}{2M_{\rm Pl}} \rho_{\rm m} \,.
\label{L3eom}
\end{equation}
This equation is of course non-linear, but nevertheless second-order --- the same amount of initial data as for an ordinary scalar field
must be supplied to obtain a unique solution. 

Since the cubic interaction term is non-renormalizable, we should treat~(\ref{L3}) as an effective field theory with cutoff $\Lambda_{\rm s}$ and allow all possible operators consistent with the symmetries\footnote{We ignore higher-order galileon terms for simplicitly. This is a technically natural choice since these terms do not get generated by quantum corrections~\shortcite{Hinterbichler:2010xn}.}
\begin{equation}
L =  -3 (\partial \pi)^2 - \frac{1}{\Lambda_{\rm s}^3} \Box\pi (\partial\pi)^2   + \sum_{n = 3}^\infty \frac{c_n}{\Lambda^{3n-4}_{\rm s}} (\Box\pi)^n  -  \frac{g}{M_{\rm Pl}} \pi \rho_{\rm m}\,,
\label{L3EFT}
\end{equation}
where the $c_n$'s are all of $O(1)$. Even if the $c_n$'s are set to zero classically, they will be generated by quantum corrections. Now, the Vainshtein mechanism~\shortcite{Vainshtein:1972sx,Deffayet:2001uk} relies on the $\Box\pi(\partial\pi)^2$ term becoming large compared to the kinetic term $(\partial\pi)^2$ near massive objects, that is, $\Box\pi \gg \Lambda_{\rm s}^3$. In this regime, one would expect that all higher-order operators $(\Box\pi)^n$ become important as well, signaling a breakdown of the effective field theory.

Contrary to this naive expectation, there is in fact a regime where the galileon term can dominate while all other operators are negligible. This is because the action~(\ref{L3EFT}) actually involves two expansion parameters~\shortcite{Luty:2003vm,Nicolis:2004qq}: a {\it classical} expansion parameter,
\begin{equation}
\alpha_{\rm cl} \equiv \frac{\Box\pi}{\Lambda_{\rm s}^3}\,,
\end{equation}
measuring the strength of classical non-linearities; and a {\it quantum} expansion parameter,
\begin{equation}
\alpha_{\rm q} \equiv \frac{\partial^2}{\Lambda_{\rm s}^2}\,,
\end{equation}
measuring the relevance of quantum corrections. We will see that there are situations where classical non-linearities are important ($\alpha_{\rm cl} \gg 1$) while
quantum corrections remain under control ($\alpha_{\rm q} \ll 1$). This is because the $(\Box\pi)^n$ operators have two derivatives per field, and hence are suppressed
by powers of $\alpha_{\rm q}$ relative to the galileon term.

Another remarkable fact is that the cubic galileon term does not get renormalized~\shortcite{Luty:2003vm,Hinterbichler:2010xn}. Indeed, all terms generated at one-loop are of the schematic form
\begin{equation}
\Gamma_{1-{\rm loop}} \sim \sum_m \left[ \Lambda^4_{\rm s} + \Lambda^2_{\rm s}\partial^2 + \log\left(\frac{\partial^2}{\Lambda_{\rm s}^2}\right) \partial^4 \right] \left(\frac{\partial\partial\pi}{\Lambda_{\rm s}^3}\right)^m\,,
\end{equation}
and hence involve 2 derivatives per $\pi$.\footnote{This traces back to the special structure of the cubic vertex. Through integration by parts, one can always arrange for an external leg $\pi_{\rm ext}$ to be hit by 2 derivatives, {\it e.g.},%
\begin{eqnarray}
\nonumber
\partial_\mu \pi \partial^\mu\pi_{\rm ext} \Box\pi &=& - \partial_\nu\partial_\mu \pi \partial^\nu\pi \partial^\mu\pi_{\rm ext} - \partial^\mu \pi \partial^\nu\pi \partial_\mu\partial_\nu \pi_{\rm ext} \\
&=& \frac{1}{2}(\partial\pi)^2 \Box\pi_{\rm ext}  - \partial^\mu \pi \partial^\nu \partial_\mu\partial_\nu \pi_{\rm ext} \,.
\end{eqnarray}
}
The galileon term, having 3 $\pi$'s and 4 $\partial$'s, is therefore not renormalized.

\section{Solution around spherically-symmetric source}

Let us illustrate this in the context of Vainshtein screening for a static point source of mass $M$, so that $T^\mu_{\;\mu} =-M\delta^{(3)}(\vec{x})$. For static, spherically-symmetric ansatz,~(\ref{L3eom}) reduces to~\shortcite{Nicolis:2004qq}
\begin{equation}
\vec{\nabla} \cdot \left(6 \vec{E} + \hat{r} \frac{4}{\Lambda_{\rm s}^3}\frac{E^2}{r}\right) = \frac{gM}{M_{\rm Pl}} \delta^{(3)}(\vec{x})\,,
\end{equation}
where
\begin{equation}
\vec{E} \equiv\vec{\nabla} \pi \,.
\end{equation}
Integrating over a sphere centered at the origin, this implies
\begin{equation}
6E + \frac{4}{\Lambda_{\rm s}^3}\frac{E^2}{r} = \frac{gM}{4\pi r^2 M_{\rm Pl}}\,.
\end{equation}
Remarkably, this equation is algebraic in $E$, and can be readily solved: 
\begin{equation}
E_\pm = \frac{\Lambda_{\rm s}^3}{r} \left(\pm \sqrt{9r^4 + 4r_{\rm V}^3r} -3r^2  \right)\,
\end{equation}
where we have introduced the Vainshtein radius
\begin{equation}
r_{\rm V} \equiv \frac{1}{\Lambda_{\rm s} } \left(\frac{gM}{4\pi M_{\rm Pl}}\right)^{1/3} = \left(g \, r_{\rm Sch}L^2\right)^{1/3}\,,
\label{rVdef}
\end{equation}
where $r_{\rm Sch} = \frac{M}{4\pi M_{\rm Pl}^2}$ is the Schwarzschild radius of the source, and where, as in~(\ref{Lsrel}), we have expressed the strong coupling scale as $\Lambda_{\rm s}  \equiv (L^{-2} M_{\rm Pl})^{1/3}$. 

This nontrivial profile is crucial to the Vainshtein effect. Below we consider two regimes. For this purpose, we will focus on the ``+" branch, with $E\rightarrow 0$ as $r\rightarrow \infty$.
(The ``$-$" branch, which asymptotically matches to the self-accelerated solution~\shortcite{Deffayet:2000uy}, has unstable ({\it i.e.}, ghost-like) perturbations~\shortcite{Luty:2003vm,Charmousis:2006pn,Gregory:2007xy}.)

\begin{figure}[t]
\centering
\includegraphics[width=5.1in]{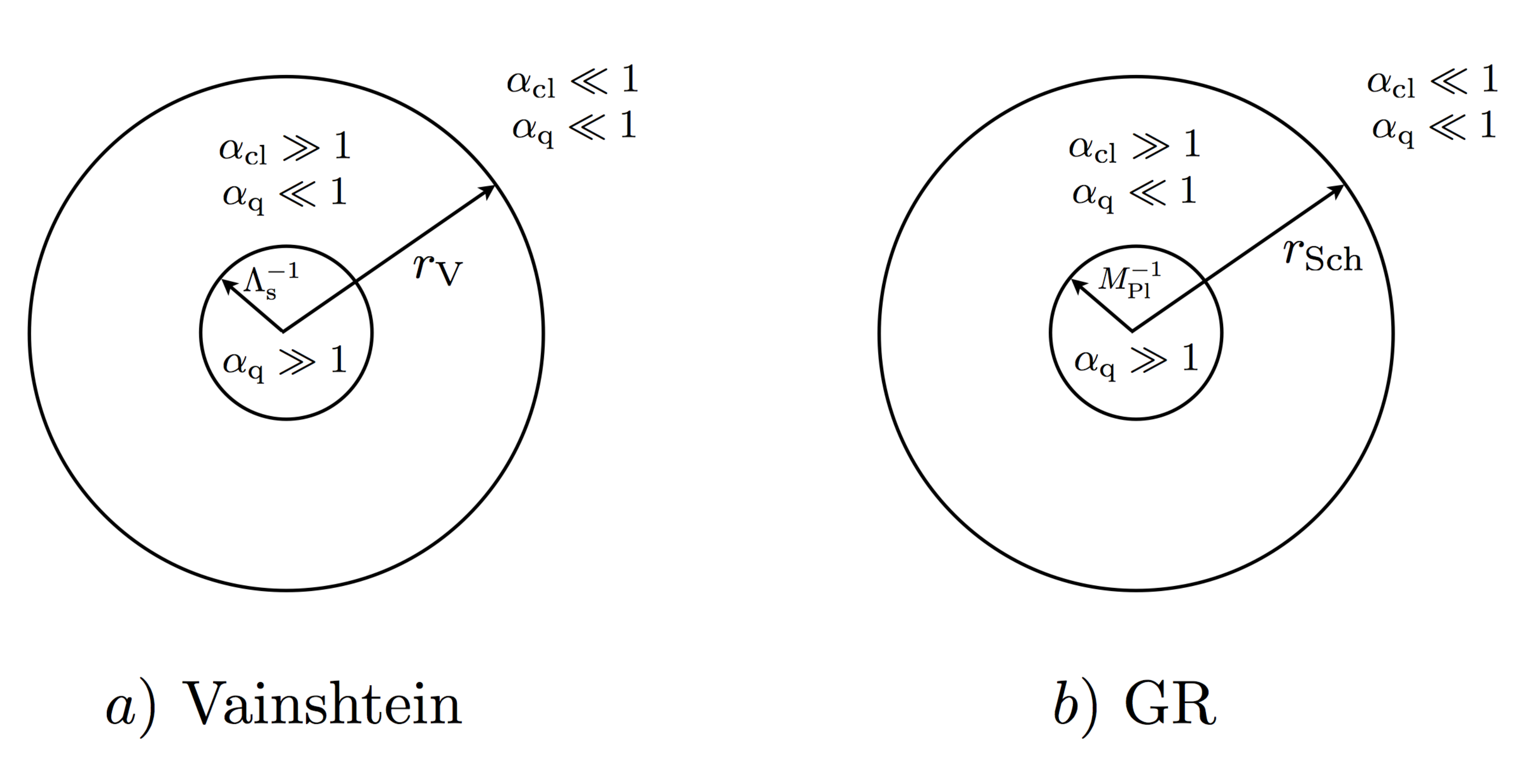}
\caption{\label{vainshtein} \small Different regimes for $a)$ Galileons and $b)$ GR. In both case, the far region ($r\gg r_{\rm V}$ for Galileons; $r \gg r_{\rm Sch}$ for GR) corresponds to the weak-field regime, where the description is both weakly-coupled and classical. In the intermediate region ($\Lambda_{\rm s}^{-1}\ll r\ll r_{\rm V}$ for Galileons; $M_{\rm Pl}^{-1} \ll r \ll r_{\rm Sch}$ for GR), the description is still classical but strongly non-linear. Below the strong coupling scale ($r \ll \Lambda_{\rm s}^{-1}$ for Galileons; $r \ll M_{\rm Pl}^{-1}$ for GR), the description becomes fully quantum.}
\end{figure}

\begin{itemize}

\item $r \gg r_{\rm V}$: Far from the source, the solution approximates a $1/r^2$ profile,
\begin{equation}
E(r \gg r_{\rm V})  =  \frac{3\Lambda_{\rm s}^3r}{4} \left( \sqrt{1 + \frac{4r_{\rm V}^3}{9 r^3} }-1 \right) \simeq \frac{g}{3} \cdot \frac{M}{8\pi M_{\rm Pl} r^2}\,.
\end{equation}
Correspondingly, the galileon force, $F_\pi = g|\nabla\pi|$, is $g^2/3$ times the gravitational force:
\begin{equation}
\left. \frac{F_{\pi}}{F_{\rm gravity}}\right\vert_{r\gg r_{\rm V}}\simeq \frac{g^2}{3}\,.
\end{equation}
(For DGP, for which $g=1$, this reproduces the famous $1/3$ enhancement.)
In this regime, both expansion parameters are small\footnote{We assume $M \gg M_{\rm Pl}$, so that $\Lambda_{\rm s}^{-1} \ll r_{\rm V}$.}:
\begin{equation}
\alpha_{\rm cl} \sim \left(\frac{r_{\rm V}}{r}\right)^3 \ll 1\,;\qquad \alpha_{\rm q} \sim \frac{1}{(r\Lambda_{\rm s})^2} \ll 1\,,
\end{equation}
hence classical non-linearities and quantum corrections are both unimportant.

\item $r \ll r_{\rm V}$: Close to the source, the solution reduces to
\begin{equation}
E \simeq \frac{\Lambda_{\rm s}^3}{2} \frac{r_{\rm V}^{3/2}}{\sqrt{r}}\sim \frac{1}{\sqrt{r}} \,.
\label{Er<<rV}
\end{equation}
The galileon force is therefore suppressed compared to gravity at distances much less than the Vainshtein radius:
\begin{equation}
\left. \frac{F_{\phi}}{F_{\rm gravity}}\right\vert_{r\ll r_{\rm V}} \sim \left(\frac{r}{r_{\rm V}}\right)^{3/2}\ll 1\,.
\label{galforcesup}
\end{equation}
In this regime, the classical non-linearity parameter is of course large,
\begin{equation}
\alpha_{\rm cl} \sim  \left(\frac{r_{\rm V}}{r}\right)^{3/2} \gg 1\,,
\end{equation}
while the quantum parameter is the same as before:
\begin{equation}
\alpha_{\rm q} \sim \frac{1}{(r\Lambda_{\rm s})^2} \,.
\end{equation}
At distances $r \gg \Lambda_{\rm s}^{-1}$, quantum corrections are under control and the classical solution can be trusted. Sufficiently close to the source,
$r \ll \Lambda_{\rm s}^{-1}$, quantum corrections become important and the effective field theory breaks down. (In fact, this conclusion is too conservative --- we will review shortly that perturbations
acquire a large kinetic term $\sim (r_{\rm V}/r)^{3/2}$, which upon canonical normalization translates to a higher effective cutoff.)

\end{itemize}
Thus, as advocated, there is a window, namely $ \Lambda_{\rm s}^{-1} \ll r \ll r_{\rm V}$, where classical non-linearities are important while quantum effects are under control. 
The various regimes are illustrated in Fig.~\ref{vainshtein}a).

This is closely analogous to what happens in GR~\shortcite{Hinterbichler:2010xn}. The Einstein-Hilbert action, expanded in powers of the canonically-normalized metric perturbation $g_{\mu\nu} =  \eta_{\mu\nu}+\frac{h_{\mu\nu}}{M_{\rm Pl}}$, is schematically of the form
\begin{equation}
L_{\rm GR} = M_{\rm Pl}^2 \sqrt{-g} R = h\partial\partial h + \sum_{n\geq 3} \frac{h^{n-1} \partial\partial h}{M_{\rm Pl}^{n-2}}\,,
\end{equation}
where we have suppressed indices for simplicity. In other words, the action consists of a kinetic term $h\partial\partial h$, and an infinite number of interaction terms with exactly two derivatives and arbitrary powers of $h/M_{\rm Pl}$. Like the galileon cubic term, the relative coefficients of these terms are not renormalized, thanks to diffeomorphism invariance. The measure of classical non-linearity is
\begin{equation}
\alpha_{\rm cl}^{\rm grav.} = \frac{h}{M_{\rm Pl}}\,.
\end{equation}
Quantum effects generate higher-curvature terms, which expanded in $h$ are of the form
\begin{equation}
L_{\rm higher-curv.} = \sqrt{-g} R^2,~\sqrt{-g}R_{\mu\nu}R^{\mu\nu} \ldots = \sum_{n\geq 2,~m\geq 4} \frac{\partial^m h^n}{M_{\rm Pl}^{m+n-4}}\,.
\end{equation}
These are suppressed relative to classical operators by powers of the factor
\begin{equation}
\alpha_{\rm q}^{\rm grav.} = {\partial^2\over M_{\rm Pl}^2}.
\end{equation}
A point source induces the spherically-symmetric profile $h\sim \frac{r_{\rm Sch}}{r}$, for which
\begin{equation}
\alpha_{\rm cl}^{\rm grav.} \sim \frac{r_{\rm Sch}}{r}\,;\qquad \alpha_{\rm q}^{\rm grav.} \sim \frac{1}{M_{\rm Pl}^2 r^2}\,.
\end{equation}
Therefore, for $r\gg r_{\rm Sch}$ (such as in the solar system), classical non-linearities are unimportant, whereas for $r\ll r_{\rm Sch}$ (such as inside and near the horizon of a black hole) they dominate. Meanwhile, quantum effects are negligible for $r\gg {1\over M_{\rm Pl}}$ but become important near and below the Planck length. The black hole horizon is the analogue of the Vainshtein radius: this is where 
classical non-linearities are large and produce important effects, while quantum effects are under control. This is illustrated in Fig.~\ref{vainshtein}b).

\section{Perturbations around the spherically-symmetric background}

Further light can be shed on the Vainshtein mechanism by considering perturbations around the spherically-symmetric background. Expanding~(\ref{L3}) in perturbations $\varphi = \phi - \bar{\phi}$
gives\footnote{Here $\partial_\Omega$ denotes the usual angular derivatives.}~\shortcite{Adams:2006sv}
\begin{eqnarray}
\nonumber
L_{\varphi} &=& \left[ 3+  \frac{2}{\Lambda_{\rm s}^3}\left(E' + \frac{2E}{r} \right)\right] \left(\dot{\varphi}^2 - (\partial_\Omega \varphi)^2\right) - \left[3 + \frac{4}{\Lambda_{\rm s}^3}\frac{E}{r} \right] (\partial_r\varphi)^2 \\
\nonumber
&-& \frac{1}{\Lambda^3_{\rm s}}\Box\varphi(\partial\varphi)^2 + \frac{g}{M_{\rm Pl}} \varphi T^\mu_{\;\mu} \\
& \sim &  \left(\frac{r_{\rm V}}{r}\right)^{3/2} \left(\dot{\varphi}^2 - (\partial_\Omega \varphi)^2 - \frac{4}{3} (\partial_r\varphi)^2 \right) - \frac{1}{\Lambda^3_{\rm s}}\Box\varphi(\partial\varphi)^2 + \frac{g}{M_{\rm Pl}} \varphi T^\mu_{\;\mu}.
\label{L3varphi}
\end{eqnarray}
where in the last step we have assumed $r\ll r_{\rm V}$ and substituted the expression~(\ref{Er<<rV}) for $E$ deep inside the Vainshtein radius.

The key point to notice is that the kinetic term is multiplied by an enhancement factor of  $\left(\frac{r_{\rm V}}{r}\right)^{3/2} \gg 1$, telling us that perturbations acquire a large inertia near a massive source.
Said differently, in terms of the canonically-normalized variable $\varphi_{\rm c} \sim \left(\frac{r_{\rm V}}{r}\right)^{3/4} \varphi$, the effective coupling to matter is reduced to
\begin{equation}
g_{\rm eff} \sim g \left(\frac{r}{r_{\rm V}}\right)^{3/4} \ll g\,,
\end{equation}
indicating that galileon perturbations decouple from matter. Moreover, the strong coupling scale $\Lambda_{\rm s}$ is renormalized to a higher scale
\begin{equation}
\Lambda_{\rm s}^{\rm eff} \sim \Lambda_{\rm s}\left(\frac{r_{\rm V}}{r}\right)^{3/4}  \gg \Lambda_{\rm s}\,,
\end{equation}
indicating that perturbations have weaker self-interactions.  

Another noticeable feature of~(\ref{L3varphi}) is that the speed of propagation in the radial direction is {\it superluminal}:
\begin{equation}
c_{\rm s}^{\rm radial} = \sqrt{\frac{4}{3}}\,.
\end{equation}
This is a generic feature of galileons --- since galileon interactions are derivative interactions, a galileon background (even an arbitrarily weak one) will deform the
light-cone for perturbations in such a way that there is always a direction along which the speed of propagation is superluminal~\shortcite{Nicolis:2009qm}. When galileons are 
coupled to Lorentz-invariant matter, this allows for the formation of closed time-like curves (CTCs). Nevertheless, it has been conjectured~\shortcite{Burrage:2011cr,Evslin:2011rj} that
galileons may be protected from the formation of CTCs by a Chronology Protection, analogously to GR~\shortcite{Hawking:1991nk}. In other words, if one tries to
create a CTC from healthy initial conditions, the galileon effective field theory will break down before the formation of the CTC~\shortcite{Burrage:2011cr}. Moreover,
it was pointed out recently~\shortcite{Creminelli:2013fxa,deRham:2013hsa} that galileon theories admitting superluminality can sometimes be mapped through field redefinitions to
healthy galileon theories, indicating that the apparent superluminalty is unphysical. This issue clearly deserves further scrutiny. 

\section{Observational Test}

The strongest bound on the cubic galileon comes from Lunary Laser Ranging (LLR) observations. (For a review, see~\shortciteN{Nordtvedt:2003pj}.) Integrated over the last 30 years, LLR monitoring constrain the Moon's orbit to $\; \lower .75ex \hbox{$\sim$} \llap{\raise .27ex \hbox{$<$}}\;  {\rm cm}$ accuracy.

Deep inside the Vainshtein radius, the galileon-mediated force given by~(\ref{galforcesup}), albeit weak, gives a small correction to the Newtonian potential:
\begin{equation}
\frac{\delta\Phi}{\Phi} \simeq \frac{g^2}{2} \left(\frac{r}{r_{\rm V}}\right)^{3/2}\,.
\end{equation}
The current constraint from LLR observations is~\shortcite{Murphy:2012rea}
\begin{equation}
\frac{\delta\Phi}{\Phi} \; \lower .75ex \hbox{$\sim$} \llap{\raise .27ex \hbox{$<$}}\; 2.4\times 10^{-11}\,.
\label{LLRcurrent}
\end{equation}
Recalling the definition of the Vainshtein radius $r_{\rm V} \equiv \left (r_{\rm Sch}L^2\right)^{1/3}$, and substituting $r_{\rm Sch} = 0.886~{\rm cm}$ for the Earth, and $r = 3.84\times 10^{10}~{\rm cm}$ for the Earth-Moon distance, the LLR constraint translates to a bound on $L$~\shortcite{Dvali:2002vf,Dvali:2007kt,Afshordi:2008rd}: 
\begin{equation}
L \;  \lower .75ex \hbox{$\sim$} \llap{\raise .27ex \hbox{$>$}} \; \frac{H_0^{-1}}{20\,g^{3/2}} \simeq 150 g^{-3/2}~{\rm Mpc}\,,
\end{equation}
where $H_0^{-1} \simeq 10^{28}~{\rm cm} \simeq 3000~{\rm Mpc}$ is the Hubble radius. The Apache Point Observatory Lunar Laser-ranging Operation (APOLLO)~\shortcite{Murphy:2012rea} is expected to improve this bound by a factor of 10. For $g\sim 1$ this would push $L$ to values larger than $H_0^{-1}$. Somewhat weaker constraints on $L$ have also been obtained (in the context of the DGP model) by studying the effect on planetary orbits~\shortcite{Battat:2008bu}.

An important property of galileons is that black holes carry no galileon hair~\shortcite{Hui:2012qt,Kaloper:2011qc}, while stars of course couple to the galileon. This leads to an interesting observational signatures: in the presence of an external linear gradient, an astrophysical black hole should be offset (by an observable amount) from the center of its host galaxy~\shortcite{Hui:2012jb}. Cosmologically, the galileon-mediated force becomes important at late times and on large scales. This affects various linear-scale observables~\shortcite{Song:2007wd,Afshordi:2008rd,Lombriser:2009xg}, such as enhanced large scale bulk flows~\shortcite{Wyman:2010jp,Khoury:2009tk}, infall velocities~\shortcite{Zu:2013joa} and weak-lensing signals~\shortcite{Wyman:2011mp}. See~\shortciteNP{Khoury:2009tk,Schmidt:2009sv,Chan:2009ew,Wyman:2013jaa} for N-body simulations.

\section{General galileons}

Beyond the cubic galileon~(\ref{L3}), there are exactly 5 independent terms in 4 dimensions which shift by a total derivative under~(\ref{galsym}), and whose equations of motion are second order.\footnote{This is a consequence of the fact that these are Wess-Zumino terms for spontaneously broken space-time symmetries~\shortcite{Goon:2012dy}.}
(More generally, there are $D+1$ galileon terms in $D$ dimensions.) In 4 dimensions, they take the form
\begin{eqnarray}  
\label{galileonterms}
L_1&=&\pi\,; \nonumber \\
L_2&=& - (\partial\pi)^2 \,;\nonumber \\
L_3&=&-(\partial \pi)^2 \Box \pi \,; \nonumber \\
L_4&=& - (\partial\pi)^2\bigg((\Box\pi)^2-(\partial_\mu\partial_\nu\pi)^2\bigg) \,; \nonumber \\
L_5&=&-(\partial\pi)^2\bigg((\Box\pi)^3+2(\partial_\mu\partial_\nu\pi)^3-3\Box\pi(\partial_\mu\partial_\nu\pi)^2\bigg)\, .
\label{gengal}
\end{eqnarray}
In other words, to the cubic action~(\ref{L3}), which comprises $L_1$, $L_2$ and $L_3$, one can also add $L_4$ and $L_5$. 
None of these terms are renormalized at any order in perturbation theory~\shortcite{Hinterbichler:2010xn}. These more general galileons arise, for instance,
in the decoupling limit of dRGT massive gravity~\shortcite{deRham:2010kj,deRham:2010kj,Hinterbichler:2011tt}.

Remarkably, the galileons form an {\it Euler hierarchy}~\shortcite{Fairlie:1992zn,Fairlie:1992nb,Fairlie:1991qe,Fairlie:2011md}. An Euler hierarchy is a sequence of Lagrangians $\{L_2,L_3,\ldots\}$ such that:

\begin{itemize}

\item $L_{n+1} =L_2 (E\,L_n)$, for all $n\geq 2$, where $E\,L_n$ denotes the Euler-Lagrange equation for $L_n$.

\item $L_n$ depends on $\partial\pi$, $\partial^2\pi$, but {\it not} on higher derivatives of the field.

\item The procedure terminates. In $D$ dimensions, one gets exactly $D$ terms in the sequence: $\{L_2,\ldots,L_{D+1}\}$.
In particular, $L_{D+2}$ is a total derivative. 

\end{itemize}

To see how~(\ref{gengal}) can be built as an Euler hierarchy, we start with $L_2 =  - (\partial\pi)^2$ and obtain
\begin{eqnarray}
\nonumber
E\,L_2 &=& \Box\pi \\
\Longrightarrow ~~~~~L_3 &=&  L_2 (E\,L_2) = -(\partial\pi)^2\Box\pi\,,
\end{eqnarray} 
which indeed agrees with $L_3$. Similarly,
\begin{eqnarray}
\nonumber
E\,L_3 &=&  (\Box\pi)^2 - (\partial_\mu\partial_\nu\pi)^2\\
 \Longrightarrow ~~~~~ L_4 &=& L_2 (E\,L_3) = - (\partial\pi)^2\bigg((\Box\pi)^2-(\partial_\mu\partial_\nu\pi)^2\bigg)\,,
\end{eqnarray} 
which correctly reproduces $L_4$, and so on.

\vspace{0.2cm}
\noindent {\bf Exercise}: Derive the Euler hierarchy starting from $L_2 = - \sqrt{1 + (\partial\pi)^2}$ to obtain the {\it DBI galileons}. See Eqs.~(38)-(41) of~\shortciteN{deRham:2010eu} for the answer.
\vspace{0.2cm}

The galileon Lagrangians~(\ref{gengal}) can be expressed succinctly as
\begin{equation}
L_{n+1} = n \eta^{\mu_1\nu_1\mu_2\nu_2\cdots\mu_{n}\nu_{n}} \partial_{\mu_1}\pi\partial_{\nu_1}\pi\partial_{\mu_2}\partial_{\nu_2}\pi\cdots\partial_{\mu_{n}}\partial_{\nu_{n}}\pi\,.
\label{galotherparameterization}
\end{equation}
The tensor $\eta$ is defined by
\begin{equation}
\eta^{\mu_1\nu_1 \cdots\mu_n\nu_n} \equiv{1\over n!}\sum_p\left(-1\right)^{p}\eta^{\mu_1p(\nu_1)}\eta^{\mu_2p(\nu_2)}\cdots\eta^{\mu_np(\nu_n)}
\end{equation}
where the sum is over all permutations of the $\nu$ indices, with $(-1)^p$ denoting the sign of the permutation. The $\eta$ tensor is 

\begin{itemize}

\item anti-symmetric in the $\mu$ indices;

\item anti-symmetric in the $\nu$ indices;

\item symmetric under $(\mu_i,\nu_i) \leftrightarrow (\mu_j,\nu_j)$.

\end{itemize}

In this form it is clear why the equations of motion are second order. Indeed, the dangerous higher-derivative terms are of the form 
\begin{equation}
E\, L_{n+1} \supset \eta^{\mu_1\nu_1 \cdots\mu_n\nu_n} \partial_{\nu_1}\pi\partial_{\mu_2}\partial_{\nu_2}\pi\cdots\partial_{\mu_1} \partial_{\mu_{n}}\partial_{\nu_{n}}\pi  \,,
\end{equation}
but this vanishes because $\eta^{\mu_1\nu_1 \cdots\mu_n\nu_n} \partial_{\mu_1} \partial_{\mu_{n}} \pi = 0$ by the anti-symmetry property of $\eta$ in the $\mu$ indices.
Hence the equation of motion involves exactly two derivatives per $\pi$, and is therefore manifestly invariant under the galilean symmetry~(\ref{galsym}).

\chapter{Summary}

This is truly an exciting time for cosmology. We have a standard cosmological model, the $\Lambda$CDM model, which thus far successfully accounts for all observational data. It is
remarkably predictive: with a single number $\Lambda$ parametrizing dark energy, the expansion and growth histories are uniquely predicted. In the next few years, the standard
model will be confronted by a host of increasingly powerful probes of the large scale structure, such as the Dark Energy Survey\footnote{{\tt https://www.darkenergysurvey.org/.}}, BigBOSS\footnote{{\tt http://cosmology.lbl.gov/BOSS/.}},  the Large Synoptic Survey Telescope\footnote{{\tt http://www.lsst.org/lsst.}} and EUCLID\footnote{{\tt http://sci.esa.int/science-e/www/area/index.cfm?fareaid=102.}}.
These precision experiments will test the $\Lambda$CDM model to unprecedented accuracy and may well reveal the existence of ``beyond-the-standard-model"
physics, in the form of new light degrees of freedom in the dark sector.

In these Lectures, we reviewed various extensions of the $\Lambda$CDM model characterized by additional scalar fields. 
In order to reproduce the successful phenomenology of GR in the solar system, these scalars must effectively decouple from matter on solar
system/laboratory scales. This can be achieved through {\it screening mechanisms}, which rely on the high density of the local environment
(relative to the mean cosmological density) to suppress deviations from standard gravity. The manifestation of the new degrees of
freedom (generally scalar fields) therefore depends sensitively on their environment, which in turn leads to striking experimental signatures. 

We presented 3 screening mechanisms, characterized by whether non-linearities are triggered by the local field
value $\phi$ (chameleon), the first derivative of the field $\partial\phi$ (k-Mouflage), or its second derivative $\partial^2\phi$ (Vainshtein). 
Observationally, the resulting modifications are most pronounced, respectively, in regions of low Newtonian potential $\Phi$ (chameleon), low acceleration
$a = |\nabla\Phi|$ (k-Mouflage), and low curvature $R = |\nabla^2\Phi|$ (Vainshtein). Chameleon effects are important on non-linear
scales today ($\lower .75ex \hbox{$\sim$} \llap{\raise .27ex \hbox{$<$}} ~{\rm Mpc}$), while k-Mouflage/Vainshtein effects are important
on linear scales today ($\lower .75ex \hbox{$\sim$} \llap{\raise .27ex \hbox{$>$}}~{\rm Mpc}$).

A tantaliziing feature of these mechanisms is that they also make testable predictions for local tests of gravity. 
The most striking signature of chameleons can be found by testing gravity in space. Because the screening condition depends on the ambient density, small bodies that are screened in the
laboratory may be unscreened in space. This leads to striking predictions for future satellite tests of gravity, such as the planned MicroSCOPE mission\footnote{{\tt http://microscope.onera.fr/}} and STE-QUEST\footnote{{\tt http://sci.esa.int/science-e/www/area/index.cfm?fareaid=127}}. In particular, chameleons can result in violations of the (weak) Equivalence Principle in orbit with $\eta \equiv \Delta a/a \gg 10^{-13}$, in blatant conflict with laboratory constraints. Similarly, the total force between unscreened particles can be $O(1)$ larger than in standard gravity, which would be interpreted as $O(1)$ deviations from the value of $G_{\rm N}$ measured on Earth.

The Vainshtein mechanism lead to small, but testable, deviations from $1/r$ gravity in the solar system. The most sensitive probe is Lunar Laser Ranging, which monitors the orbit of the Moon to incredible accuracy. The ongoing APOLLO experiment has the sensitivity to place interesting constraints on the simplest galileon theories. Specifically, it can probe galileons whose characteristic scale $L$ is
of order of the present Hubble radius.

\acknowledgements{I wish to warmly thank the organizers of the 2013 Post-Planck Cosmology Summer School for inviting me to lecture at Les Houches,
and all the students at the School for your enthusiasm and many stimulating discussions. 
%Special thanks to Paolo Creminelli, Simone Ferraro, Gabriele Trevisan, 
This work was supported in part by NSF CAREER Award PHY-1145525, NASA ATP grant NNX11AI95G, and the Alfred P. Sloan Foundation.}

%\thebibliography{0}
\bibliography{KhouryLesHouchesarchive}{}
\bibliographystyle{OUPnamed_notitle}

\end{document}